\documentclass[12pt]{article}
\usepackage[dvips]{graphics,color}
\def\bea{\begin{eqnarray}}
\def\eea{\end{eqnarray}}
\def\be{\begin{equation}}
\def\ee{\end{equation}}
\def\nn{\nonumber}
\def\le{\left}
\def\re{\right}

\def\i{\imath}

\def\a{\alpha}
\def\b{\beta}

\def\d{\delta}

\def\e{\epsilon}

\def\g{\gamma}
\def\G{\Gamma}

\def\l{\lambda}

\def\m{\mu}
\def\n{\nu}

\def\r{\rho}
\def\s{\sigma}
\def\t{\tau}

\def\x{\xi}
\def\z{\zeta}

\def\p{\partial}

\begin{document}
\begin{flushright}
{\small AEI-2002-011}
\end{flushright}

\begin{center}
{\large \bf {The {\it Real} Wick rotations in quantum gravity} 
}

\vspace{2cm}

Arundhati Dasgupta\footnote{{\it email}: dasgupta@aei-potsdam.mpg.de}, \\

Max-Planck-Institut f\"ur Gravitationsphysik,\\
Am M\"uhlenberg 1, D-14476 Golm, Germany.

\end{center}

\vspace{3cm}

We discuss Wick rotations in the context of gravity, 
with emphasis on a non-perturbative Wick rotation proposed in hep-th/0103186  
mapping real Lorentzian metrics to real Euclidean metrics in
proper-time coordinates. As an application, we demonstrate how this Wick 
rotation leads to a correct answer for a two dimensional 
non-perturbative path-integral.

\newpage

\section{Introduction}
The evaluation of the path-integral for any field theory, essentially
involves a rotation to Euclidean signature space-times. This renders the
complex measures and oscillatory weight factors $ {\cal D}g e^{\i S}$ real and the functional
integral computable. One then Wick rotates back in the final
answer to get Lorentzian quantities, or invoke the Osterwalder-Schraeder
theorems. The usual methods used for field theories in flat space-times are difficult to extend
to gravity as there is no distinguished notion of time. There
have been attempts to define a Euclidean gravitational path-integral
and relate them to Lorentzian quantities. Initially complex weights were replaced
by real Boltzmann weights and the sum performed over all possible Riemannian
metrics \cite{haw}. This prescription is ad hoc as there is no a priori relation 
between the Lorentzian and Riemannian sums. However, if one believes
that Lorentzian quantities are the relevant physical quantities, then
a prescription must exist for identifying the corresponding Euclidean metrics
to be summed over. The usual Wick rotations pertain to static metric solutions
of Einstein's equations. Given the time coordinate, $t$, Wick rotation
amounts to going to the space of complex metrics, and integrating
over the imaginary time axis. As evident, this prescription cannot be directly 
extended to a non-perturbative configuration space path integral where
there is a priori no distinguished notion of time.

 The solution to this problem was proposed in \cite{dalo}, where one gauge fixes the metric and performs a 
Wick rotation on the space of gauge fixed metrics. 
This `non-perturbative' Wick rotation is a first step in mapping Lorentzian path-integrals to Euclidean path-integrals, in configuration space. The mapping crucially does not involve 
a complexification of space-time and maps the real Lorentzian metrics to corresponding real Riemannian metrics in the path-integral. To verify that our proposal works, we discuss some examples where an explicit answer can be obtained. In particular we show how in 2 dimension, the only case 
where the `non-perturbative' sector can be explicitly evaluated, a correct 2-loop answer is obtained.
We show how the rotation relates the Lorentzian and Euclidean sectors
of the theory. A completely independent evaluation of the Euclidean path integral in 2 dimensions would include `baby universes' and hence not satisfy
the stringent causality requirements which the Lorentzian histories have 
to satisfy. We first define the path-integral in Lorentzian space, gauge fix the path-integral, Wick
rotate to Euclidean space to perform the computations. We also restrict
the topology of the manifold to be fixed. The Faddeev-Popov determinant which arises due to gauge
fixing gives the effective action in proper-time coordinates with
the only degree of freedom being $g_{11}$ component of the metric.
Using a heuristic argument by coupling 2 dimensional gravity to matter, we show that this component of the
metric has the product form $g_{11}= \g(t)\g(x)$. This immediately reduces
the problem to that of quantum mechanics, and the zeta function
regularisation of the Faddeev-Popov determinant gives the
effective action as a quantum mechanical action. The path-integral
can be explicitly evaluated once this is determined. We then 
rotate the final result to obtain the physical Lorentzian answer.
The answer obtained in the end is same as that obtained in a discretized
evaluation of the path-integral \cite{amlo}. A continuum calculation
of the path-integral in proper-time coordinates had been
previously obtained in \cite{nak}. Our approach is along the formalism of \cite{dalo}, where we derive the effective action from the Faddeev-Popov determinant arising in the measure unlike \cite{nak}. {\it Note the choice of gauge the Proper-time gauge plays a crucial role
in all our arguments on the Wick Rotation as Lorentzian concepts like causality are manifest, which restricts the path-integral.}

The following section is a detailed discussion of the Wick rotation proposed in \cite{dalo}, and the reason
for it being non-perturbative is pointed out. The discussions in this section are similar as those in \cite{dalo2}. The third section enumerates the applications. An explicit
calculation of the two dimensional path-integral is given, and there is a short review of the conformal mode
cancellation of \cite{dalo} to complete our argument. We end with a discussion.
\section{The Non-perturbative `Wick Rotation'}
Since the {\it perturbative} path integral does not define
a fundamental theory of quantum gravity, we have to 
consider, non-perturbative evaluations (see \cite{ajl} for
 recent examples). As there is no distinguished notion of time in gravity, definition
of Euclidean path-integral is not well defined.  A non-perturbative Wick rotation however can be 
implemented on the space of gauge fixed metrics as shown in \cite{dalo}.
Here, after
fixing the gauge in the {\it Lorentzian} path integral, 
the appropriate gauge-invariant metrics are written in coordinates with a distinguished time . We then `Wick rotate' in the
time-like component of the gauge invariant metric. To implement
this, we need to identify a suitable gauge, and the proper-time
gauge precisely fits in. Here the notion of time is distinguished, and 
causality is manifest.
Our construction, including the `Wick rotation', is non-perturbative 
and background-independent.

Our gauge is defined by the use of Gaussian normal coordinates, 
where the $d$-dimensional metric takes the form
\begin{equation}
ds^2= \epsilon \ dt^2 + g_{ij} dx^idx^j,\;\; \epsilon=\pm1,\;\;
i=1,2,...,d-1,
\label{metric}
\end{equation}
singling out a {\it proper time} $t$. The value  $\epsilon=-1$
corresponds to Lorentzian-signature metrics.
Our generalized Wick rotation consists in 
mapping 
\begin{equation}
\epsilon\mapsto -\epsilon,
\label{wick2}
\end{equation}
while keeping $g_{ij}$ {\it unchanged}. 
It has the following properties:

\includegraphics{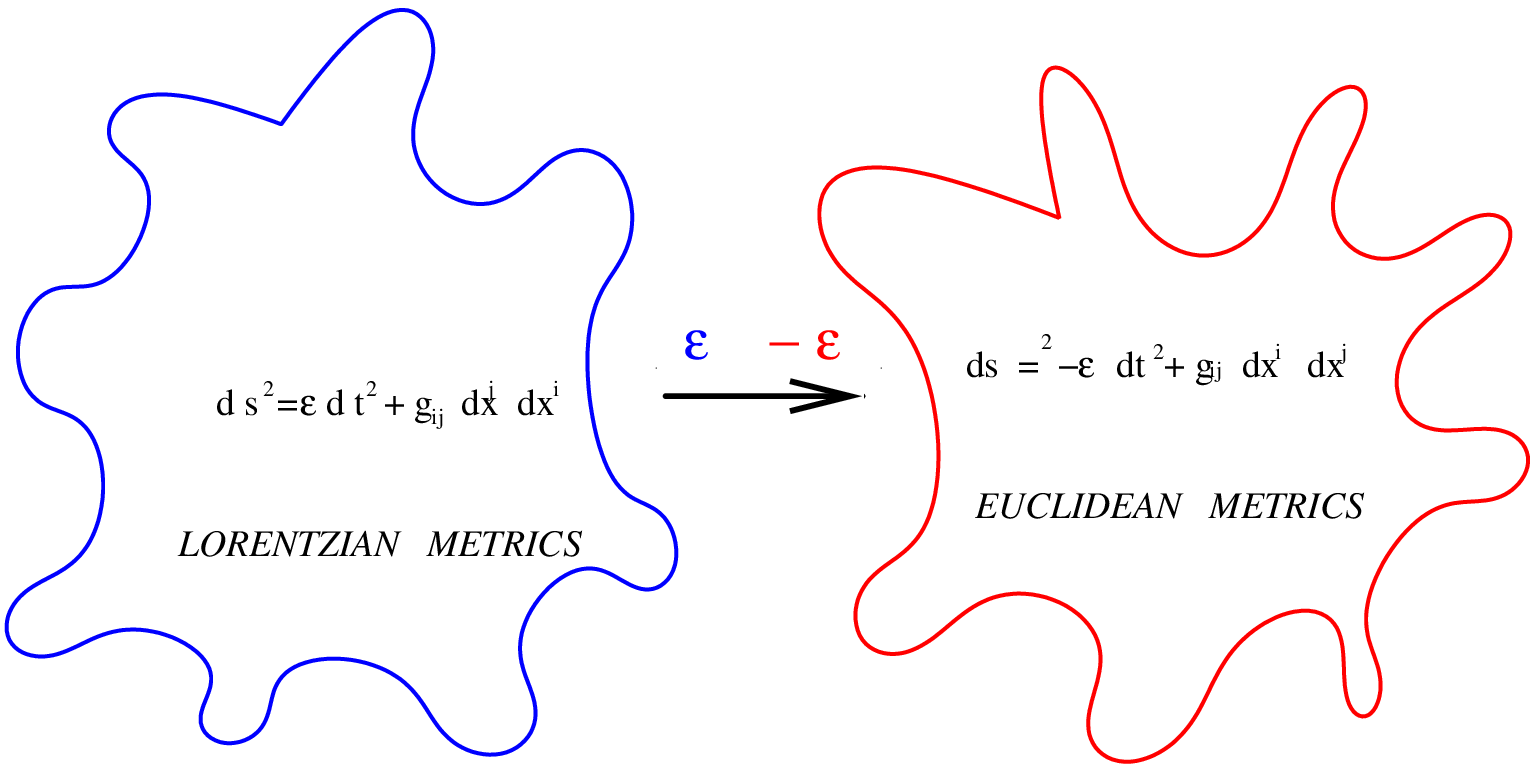}

\begin{itemize}
\item{ {\it Real} Lorentzian metrics are mapped to {\it real}
Euclidean metrics, thus making the DeWitt metric
and associated scalar products well-defined. Note that this is unlike the 
prescriptions used in \cite{york} where complexification of the metric 
occurs. This is necessary for us as we need not only the action to be real
but also the De-Witt metric $G^{\mu \nu \rho \sigma}$ in the configuration space of metrics, and the
associated scalar products $<h, h> = \int d^4x \sqrt{g} h_{\mu \nu} G^{\mu \nu \rho \sigma} h_{\r \s}$ to be real too.
A simple illustrative example is that of
the de Sitter metric (foliated by flat hypersurfaces), 
\begin{equation}
ds^2= \epsilon \ dt^2 + e^{2\Lambda t} \  d\vec x^2,
\end{equation}
which is also a solution to the Einstein equations. 
The standard rotation $t\rightarrow it$ takes this to a complex-Euclidean
metric. Not only that, the scalar product defined above is clearly 
complex for d=4. The Gaussian normalisation conditions used to define
the correct measure in the form
\be
\int {\cal D} h \exp^{- <h,h>}=1
\ee
will no longer have a real exponent, and hence the `Euclideanised' integrals
will also be ill defined.
 The Wick rotation to real Riemannian metrics which is achieved in (\ref{wick2})
is a necessity here.}
\item{ The initial motivation for Wick rotation is of course the replacement of
complex amplitudes in the path-integral by real weights. This is very well
achieved in our prescription. The Complex amplitudes which appears in the path integral can be written in this gauge as ${\rm e}^{i\sqrt{-\epsilon} 
S_{\epsilon}}|_{\epsilon =-1}$, with the action S as:
\begin{equation}
S_{\epsilon}[g]=
-\frac{\epsilon}{16\pi G}\int d^dx \sqrt{\det g_{ij}}
\left({}^{(d-1)}R -2\Lambda - \frac{\epsilon}{4}
G^{ijkl}_{(-2)} (\partial_0 g_{ij})(\partial_0 g_{kl})
\right),
\label{actfix}
\end{equation}
where $G^{ijkl}= (g^{ik}g^{jl} + g^{il}g^{jk} + C g^{ij}g^{kl})/2$ 
is the DeWitt metric defined on the
functional space of metrics and $\Lambda$ the cosmological constant. 
(The one-parameter ambiguity $C$ gets fixed
by the form of the kinetic term in the action to $C=-2$.)
Now clearly, the Wick rotation prescription given as $\e\rightarrow -\e$
helps in mapping the complex weights to real Boltzmann weights
 ${\rm e}^{-S_{eu}}$.}
\item{ The map is one-to-one and onto the space of gauge-fixed Euclidean 
metrics. We assume that metrics of the form (\ref{metric}) 
are in a suitable sense dense in the quotient of metrics modulo
diffeomorphisms, see \cite{dalo} for a detailed discussion.} 
\item{ The non-perturbative nature of our Wick rotation becomes manifest when we look at {\it solutions} to Einstein's equations. In general, Lorentzian solutions of the Einstein equations
are not mapped to solutions of the Einstein equations in Euclidean
space by (\ref{wick2}). In proper-time coordinates, as written in (\ref{metric}), the entire
$\epsilon$ dependence can be absorbed into a rescaling of the time
coordinate by $\sqrt {\e}$ (See Appendix A). Thus Lorentzian and Euclidean
solutions  ($g_{ij}^L (\t,r) = g_{ij}^E(\t, r)$) have the same functional dependence on the scaled parameter $\t= \sqrt{\e} t$, however, $\e \rightarrow -\e$
would implement the change  $t \rightarrow \i t$. Note that in our proposal
for the Wick rotation, we assume that $g_{ij}$ is independent of $\epsilon$.
Thus we are essentially not looking at Classical metrics, or Saddle points
of the path-integral. The usual `common-sense' associated with Einstein's
equation solutions cannot be implemented in our space of metrics.}
\item{Another positive point of this Wick rotation, is the fact that since
$g_{ij} (t)$ remains unchanged, whatever properties the Lorentzian metric has
as a function of Lorentzian time are retained in the Euclidean case.
As an example, the metric for desitter space (foliated by spheres, and $\Lambda=1$) is:

\be
ds^2 = - dt^2 + \cosh^2t (d \chi^2 + \sin^2\chi d\Omega^2)
\label{dem}
\ee
 The standard $t\rightarrow it$ would give, 
\be
ds^2 =  dt^2 + \cos^2 t (d \chi^2 + \sin^2 \chi d\Omega^2)
\label{deme}
\ee
which is pathological at $t=\pm \pi/2$.

However, our Wick rotation gives  
\be
ds^2 =  dt^2 + \cosh^2 t (d \chi^2 + \sin^2\chi d\Omega^2)
\label{demep}
\ee

with the result that the
Euclidean metric is not pathological at $t=  \pm \pi/2$. Hence, once we have
fixed the configuration space of metrics for the Lorentzian case, we can retain
the mapped Euclidean set of metrics in the sum.}
\end{itemize}

The prescription (\ref{wick2}) and the $t\rightarrow it$ agree only 
in a few cases, for example, the Minkowski
and Schwarzschild metrics (which are static). 
If we were primarily interested in a perturbative treatment
around given classical solutions,
we might worry about the second last property. However, we need a
Wick rotation that works on the space of
{\it all} gauge fixed metrics, since this is the space we wish to sum over
in the non-perturbative path integral, for given initial and final
boundaries.

We conclude that in a non-perturbative context the prescription
$\epsilon\mapsto -\epsilon$ is perfectly well-defined, and
clearly preferred to a rotation to imaginary (proper) time. 

\section{Applications}
The Wick rotation discussed in the previous section has to be implemented
in a actual path integral to illustrate that it actually works. 2 dimensions is simplest
to start with, and a full non-perturbative path-integral can be evaluated. 
\subsection{PI in 2 dim} 
The exact integral in 2 dimensions, helps
in illustrating our Wick rotation, which is specific to 
the proper-time gauge we have chosen. We define the
path-integral in two dimensional Lorentzian geometry before
`Wick rotating' to Euclidean space-times. 
The  choice of our gauge helps us in realising Lorentzian conditions
manifestly e.g.\\
1)The existence of a global proper-time coordinate ensures that Topology is
restricted (eg: formation of baby Universe would require a degenerate metric
at the point of splitting, with the light cones pointing towards different
time directions subsequently. See for reference \cite{loll,dow}.)\\
2)Since propagator is evaluated between proper-time slices with the
proper-time flowing from $[0,T]$, causality is manifest in this gauge.\\ 

The path-integral is
\be
Z= \int {\cal D}{g_{\mu\nu}^{\rm P}}~ J ~\exp(- \i S)
\ee
where J is the Faddeev-Popov determinant (FP) which arises due to the gauge fixing
to a metric ${g_{\m \n}^{\rm P}}$. After determining the dependence of the
FP on the metric variable (which in two dimensions has only one free
component), one proceeds to evaluate the entire path-integral. (Here we have removed the diffeomorphism degrees of freedom.) 

The gauge fixed metric $g^{\rm P}_{\m \n}(t,x)$ has the following form:
\be
ds^2 =  \e dt^2 + \g(t,x) dx^2,
\label{ptm}
\ee

with $\g(t,x)$ being the $g_{11}$ component of the metric. 
Before we start with evaluation of the P.I., 
we look at the classical phase space in proper-time coordinates.
Given the two dimensional classical action $S= \int \sqrt{g} R + \Lambda\int \sqrt g$, ($\Lambda$ is the cosmological constant),
one has to impose the constraint $T_{01}=0$,
to determine the classical configuration space(Appendix B). Imposing this condition reduces the configuration space such that
 $\partial_t\partial_x \ln \sqrt \g(t,x)=0$,
and hence $\g =\g(t)\g(x)$. With the above variables, the problem reduces to one of quantum mechanics. 
This restriction, is also closely related to the 
restriction to cylinder topology.
As seen in discrete calculations \cite{amlo}, one essentially studies
correlations of loops $\int \sqrt{\g} dx=l(t)$ in time, with the space direction integrated out. (One can attribute this to Lorentzian cobordism which restricts possible classical geometries between two given boundary data in 2 dimensions. For discussions on these issues see \cite{dow,vis}). 

\begin{center}
\includegraphics{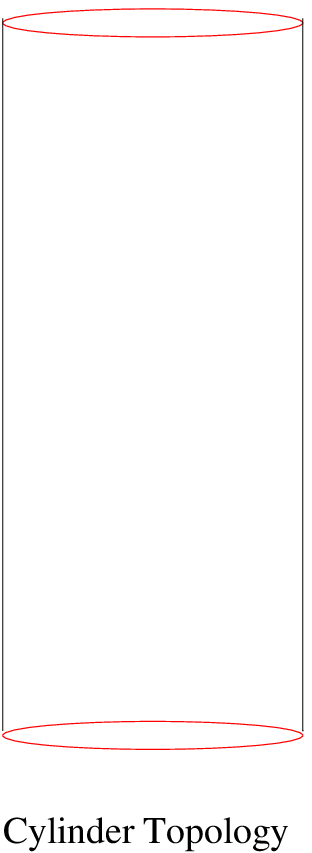}
\end{center}

It is an interesting exsersice to determine the behaviour of
 2-dim gravity coupled to matter fields in this gauge. Using
a heuristic argument, we demonstrate how quantisation in this
gauge is different from a ab initio quantisation in conformal gauge. 

As explained in \cite{poly} the matter integral can be done
exactly, and the effective energy momentum tensors determined.  
In other words: $<T_{\m \n}> = \frac{\d F}{\d g^{\m \n}}$,  with

\bea
Z^{M}= \int {\cal D}{X} \exp(-\frac12\int{\sqrt g} g^{\m \n} \partial_\m X\partial_\n X) &= &\exp( -F) 
\eea

The conformal anomaly 
(in Euclidean space) is given as \cite{poly}

\be
g^{\m \n }<T_{\m \n}> = (\frac{D}{2 4 \pi}) \le[ R(\xi) + constant\re]
\label{conf}
\ee

The $D$ denotes the number of scalar fields.
Note, the presence of the conformal anomaly does not break the diffeomorphism
invariance of the quantum effective action. Now, we impose the constraint
in proper-time gauge such that $<T_{01}>=0$. This can be implemented in the 
effective action by using the following property:
\be
 g_{01} = h_{01}
\ee
such that
\be
\frac{\d Z^M}{\d h_{0 1}}|_{h_{0 1}=0} = <T_{0 1}>= 0
\ee

This restriction in proper-time gauge immediately implies that
the correlations are important only across two proper-time
slices. 
 Let us try to see what this implies for conformal gauge.
The transformations which take the conformal set of coordinates
to the proper-time coordinates are as given below.
The metric in conformal gauge can be written in terms of $(\xi_0, \xi_1)$ coordinates as the following:
\be
ds^2 = e^{2\phi(\xi)}\le(\e d\xi_0^2 + d\x_1^2\re) \label{conm}
\ee

Where $\phi$ denotes
the conformal mode and the
$\epsilon \rightarrow -\epsilon$ denotes a corresponding mapping to
Euclidean metrics in conformal gauge. This mapping is of course not
equivalent to the mapping defined in the proper-time gauge. For simplicity
at present we take the $\e$ to be same.
The coordinate transformations are very simple, and gives a relation
between the derivatives of the coordinates as (see Appendix B):
\bea
\e ~\dot t ^2 + t^{' 2} &= & \e~ e^{2\phi} \label{trans1} \\
\e \dot x^2 + x^{' 2} &=& \frac{\exp(2\phi)}{\g} \label{trans2}\\
{\mbox {with}}&& \dot t= \pm \sqrt{\g} x', \ \ \ t'= \pm \sqrt{\g} \dot x
\label{trans3}
\eea

(Where $\dot t$ denotes differentiation of t with respect to
$\xi_0$ and $x'$ denotes differentiation of $x$ with respect 
to $\xi_1$).
This coordinate transformation can be implemented on the energy
momentum tensors in (\ref{conf}). 

To quantise in conformal gauge, one imposes the constraint $<T_{01}^{CF}>=0$, and one can use the diffeomorphism transformation to see what it
implies in the proper-time coordinates. Using the coordinate transformations,
one finds:
\be
<T_{01}>^{CF}= ~\dot x ~x'~\left(\e <T_{00}> - \frac{1}{\g} <T_{11}>\right)=0 
\ee
Where $<T_{01}>=0$ has been implemented. Clearly for the above to be correct,
 one is left with the following
options:
\bea
\dot x=0,& {\rm or}&  \ x'=0 \ \ \  {\rm or} \ \\ 
\e <T_{00}> &= &\frac{1}{\g}<T_{11}>
\label{cond}
\eea
 
The last condition is disallowed from equations which arise determining
the energy momentum tensors due to (\ref{conf}) and the conservation equations
$\nabla^{\mu}<T_{\mu \nu}>=0$ (Appendix B). 
 
It also happens that when we continue to Lorentzian signature by 
(\ref{trans2}),the condition $x'=0$ is not possible.  
 What we have left is $\dot x=0$ and hence
by (\ref{trans3}),  $t'=0$, which implies again that
$\g= \g(x)\g(t)$ and $\phi(\xi)=\phi(\xi_0)$.  
Note in principle we should have used the inverse transformations,
i.e. transformed $<T_{01}>=0$ to conformal gauge, but conclusions 
would have been same.

An ab initio quantisation in conformal gauge would not have the
reduced the conformal mode to be time dependent, and yielded the same result.
{\it Thus our choice of gauge on which we can implement the Lorentzian conditions in a manifest manner, is crucial.}

The path integral takes the following form in the proper-time gauge:
\be
Z= \int {\cal D}\g J(\e,\g) e^{i \sqrt{\e} \Lambda \int d^2x \sqrt{\g}}.
\label{act}
\ee
 
The Wick rotation which takes $\e\rightarrow -\e$
is very simple , and we compute the entire path integral calculation in Euclidean space. {\it Albeit hereafter the computations are similar to a Euclidean
P.I, one must realise that an ab initio calculation in Conformal gauge 
Euclidean P. I. would not have resulted in the same answer, unless one
implemented the Lorentzian conditions and rotated the corresponding proper-time
obtained as a function of the conformal mode}.

The main task is the evaluation of the Faddeev-Popov determinant J and it's dependence
on $\g$. To evaluate the determinant explicitly, we refer to \cite{dalo}
and use the Gaussian normalisation condition to fix the exact value.
We expand the metric around the diffeomorphism orbits:
\begin{equation}
\delta g_{\mu \nu} = \delta g_{\mu \nu}^{\rm PT} + \nabla_{\mu}\xi_{\nu}
+ \nabla_{\nu}\xi_{\mu} = \delta g_{\mu \nu}^{\rm PT} + (L\xi)_{\m \n}
\label{diff}
\end{equation}

For our purposes we set C=0, a value predetermined by the action we have chosen.
(Note that this is very specific to 2-dimensions)

(\ref{diff}) is then used in:
$$\int {\cal D} \delta g_{\mu \n} \exp(-<\delta g,\delta g>)=1$$
 Denoting the
Jacobian of the transformation as $J$, the induced measure on the tangent
space is determined. Note that since the exponent is diffeomorphism 
invariant, we determine the scalar product at the base point on the
gauge slice. As given in Appendix C, the Faddeev-Popov determinant
 can be written as a product of three scalar determinants. This follows as after a few
redefinitions as in Appendix C, the entire exponent can be written as
 a sum of complete squares. 
\be
\int {\cal D} \d g_{11} {\cal D} \xi_0{\cal D} \xi_1 \exp[- \int d^2 x\le(
\d g_{1 1}(\nabla_{11})\d g_{11} + \xi'_0(\nabla_0)\xi'_0 + \xi'_1(\nabla_1)\xi'_1\re)]
\ee
Where the vectors have been re-defined in order to complete the squares as given
in Appendix C. The Gaussian integrals can now be done to yield the appropriate
Jacobians in tangent space. The Jacobians evaluated here also determine the
measure in base functional space. 
The Faddeev-Popov determinant is obtained as a product of three scalar
determinants.
 
\be
J ^2 = det_{\rm S}(\nabla_0)det_{\rm S}(\nabla_1) det_{\rm S}(\nabla_{11})
\ee
where the explicit expressions for the $\nabla_0, \nabla_1, \nabla_{11}$ are given as in Appendix C.
Being the simplest, we concentrate on $\nabla_0$ first.

\be
det_{\rm S} (\nabla_0) = det_{\rm S} \le[4(\partial^{\dag}_0\partial_0 + \frac{1}{2\g}\partial^{\dag}_1\partial_1 + \frac{(\dot{ln \g})^2}{4})\re]
\ee

The operator in the above is thus:
\be
P= \sqrt{\g}\le[4(\p^{\dag}_0\partial_0 + \frac{1}{2\g}\partial^{\dag}_1\partial_1 + \frac{(\dot{ln \g})^2 }{4})\re]
\d(t-t')\d(x-x')
\ee

The determinant of this operator then satisfies the following eigenvalue equation:
\be
P \psi = n \sqrt{\g} \psi
\ee 
where the eigenfunctions $\psi$ with eigenvalues n, have to satisfy appropriate boundary conditions on $t=0$ and $t=T$ surfaces.
We impose Dirichlet boundary conditions, and as shown below, assume that the operator equation
can be solved by a separation of variables. Once the separable solutions are found, any
other solution can be expressed as a linear combination of them \cite{mf}. To achieve
separability of the solution, we write the operator equation in the following form
\be
4\frac{\g(t)}{\psi(t)}\le(\p^{\dag}_0\p_0 + \frac{(\dot{ln \g})^2}{4}\re)\psi(t) - n \g(t) = \frac{1}{2\psi(x)\g(x)}\le(\p^{\dag}_1\p_1\re) \psi(x).
\label{sep}
\ee

What is evident from the above is the fact that the above equation is easily separable in the variables
but with the consequence that the eigenvalue $n$ depends on the time dependent operators only. This is a clear indication
of the reduction of the system to a quantum mechanical system with the reduced action
depending only on the time coordinate. (The effective action is derived from 
$J = \Pi n$. Also the
splitting we found in $\g=\g(t)\g(x)$ contributes crucially to our conclusion.)
 The eigenvalue equation for the $t$ dependent part becomes now
\be
4\g(t)\le(-(\p_0 + \frac12 \dot{\ln\g} ) \p_0 \re)\psi(t) + \g(t)(\dot{\ln\g})^2 \psi(t) - n \g(t) \psi(t) = -\a \psi(t)
\ee

where $\a$ is a positive constant independent of $t,x$. 

Finding the effective action hence reduces to the problem of finding the
determinant of the operator
\be
4 (\p^{\dag}_0\p_0) + (\dot{\ln\g})^2 + \frac{\a}{\g}.
\label{op}
\ee

In order to determine the above, we perform a zeta function regularisation.
We follow \cite{ram} to perform this. Since it is difficult to evaluate the
exact determinant, we try to guess the dependence on the $\g(t)$, by
assuming that the effective action has the form 
\be
\G= \int d t \le(f(\g)(\p_0\g)^2 + V(\g)\re)
\label{effa}
\ee
with $f(\g)$ an arbitrary function.
Thus, for 
 configurations which have $\dot{\ln \g}=0$, we recover $V(\g_0)$, where
$V(\g_0)$ is the potential depending on the constant value $\g_0$. In other words the potential is simply given by the zeta function regularisation of the
operator in (\ref{op}) with the value of $\g_0$:
\be
\int d t V(\g_0) =  -\z'(0)  
\label{ape}
\ee

Note that the
metric in (\ref{ptm}) is still non-constant due to the dependence on $\g(x)$.

Thus, we consider the operator for $\g_0$ which reduces to the following form
\be
 -4\p_0^2 + \frac{\a}{\g_0}  
\label{opc}
\ee
The Heat Kernel for this operator satisfies the following equation:
\be
-(4\p_0^2 - \frac{\a}{\g_0})G(t, t', \t)= - \p_{\t}G
\ee

with the boundary condition that
\be
G(t,t',\t=0)= \frac1{\sqrt \g_0}\d(t -t')
\ee
(the metric in this direction is flat) 
The heat Kernel $G(t, t',\t)$ can be solved to give:
\be
G(t,t',\t) = \sqrt{\frac{\pi}{4\t\g_0}} e^{-\frac{(t - t')^2}{16 \t} - \frac{\a}{\g_0}\t}
\ee

The $\z$ function is thus:
\be
\z(s) = \frac{1}{\G(s)} \int_0^{\infty} d\t \t^{s -1} \int d t{\sqrt{\frac{\pi}{4\t}}} e^{-\frac{\a}{\g_0}\t}
\ee

The $\int d t$ is not infinite,
and cancels with the volume element in the LHS of (\ref{ape}). 
The task thus reduces to the determination of
\be
\z'(0) = \frac{\sqrt{\pi}}{2}\le. \frac{\G(s-1/2)}{\G(s)}\le (\frac{\g_0}{\a}\re)^{s- 1/2}\le[ \ln\le(\frac{\g_0}{\a}\re)  + \Psi(s-1/2) - \Psi(s) \re] \re|_{s=0}
\ee

At this juncture, we keep only finite terms
as $s\rightarrow 0$, and hence only the term proportional to $\Psi(s)/\G(s)$  of the above
expression. It is easily justifiable as the rest of the terms, due to the pole of $1/\G(s)$, 
at $s\rightarrow 0$ contribute vanishingly to $ e^{-\zeta'(0)}$.

The potential
has a dependence of the form:
\be
\int d t V(\g_0) \approx \int d t \sqrt{\g_0}\frac{1}{\g_0}
\ee

Next we look at the operator $\nabla_{1}$, which has the form as given in the Appendix:
\be
\nabla_1 = \frac{4}{\g^2}\nabla^0_1\left( 1 + (\frac{\nabla^0_1}{\g})^{-1}{\nabla^1_1}\right) 
\label{ope}
\ee

with
\be
\nabla^0_1= \le(\partial_1^{\dag}\partial_1 + \frac{\g}{2}\partial_0^{\dag}\partial_0 + \frac{\g}{2}(\dot{\ln \g})\partial_0 
+ \frac{(\ln \g')^2}{4}\re)
\label{ope1}
\ee
The operator in (\ref{ope1}) can be cast in a form similar to the operator $\nabla_0$ with a  
measure of $1/{\sqrt\g_0}$. However, this does not modify the final answer, and one gets a very similar analyses as above, except for a change of factor 2. 
And the contribution to the effective potential is of the form $\int d t 1/\sqrt{\g_0}$. The rest of the factorised operator in (\ref{ope}) does not contribute
to the potential (see Appendix C). The full form of the potential {\it is} 
determined above, the contributions due to nonzero $\dot{\ln \g}$, will contribute to the derivative terms in the action. Though one may think that a constant
$\g$ can be absorbed into a rescaling of the coordinate, note that the conformal anomaly precisely prevents this.
To determine the derivative term in (\ref{effa}), we resort to the demand of locality of the
action as previous work in other gauges \cite{poly} have local actions. To see what such a term in the action should look like we make a scaling of just the 
$t$ coordinate to go to conformal gauge with the metric $$ ds^2 = \g(~d t'^2 + dx^2).$$ The
local term in these set of coordinates would be $\int dt' \frac14(\partial_t'\ln \sqrt\g)^2$. Translating this back to original coordinates we conclude that the effective action can be written as:

\be
I = \int d t\le( \frac1{4 \sqrt{\g(t)}}\le(\frac{d {\sqrt\g(t)}}{dt}\re)^2 + \frac{\b}{\sqrt \g(t)} + \Lambda \sqrt \g(t)\re)
\label{eff}
\ee
The constant $\b$ is finite.
Remarkably, the above coincides with the effective action found in \cite{nak}. To extract the finite part of our effective action, the observation of separability proved crucial. In principle
 we have not argued for the absence of higher derivative terms, however conditions of locality
are stringent. A full determination of the determinant for nonzero $\dot{\ln \g}$ should ultimately give (\ref{eff}), and shall be interesting to evaluate.

Once the action has been determined, one needs to complete the evaluation of the propagator
\be
G(l_1,l_2,T) = \int_{l_1}^{l_2} \frac{{\cal D}{l}}{l} exp(- I(l))
\label{pi}
\ee
where we are now using the variable $\sqrt{\g(t)}= l$. Though the action is quantum mechanical,
it is not clear that the evaluation of (\ref{pi}) is simple. An exact evaluation is left
for a future publication. Instead, one resorts to the following representation of the
propagator.
\be
G(l_1,l_2, T)= <l_1| e^{- HT}|l_2>
\label{ham}
\ee 
Where $H$ is the Hamiltonian corresponding to (\ref{eff}).
This is similar to the approach in \cite{nak}, and one refers the readers
to the calculations there, and quotes the final result.
\be
G(l_1,l_2,T)= \sqrt{\Lambda} e^{- \sqrt{\Lambda}(l_1 + l_2) \coth(\sqrt \Lambda T)}\frac{1}{\sinh(\sqrt \Lambda T)} I_a\le(\frac{2\sqrt{\Lambda l_1 l_2}} {\sinh( \sqrt\Lambda T)}\re)
\ee

This amplitude is the same obtained in different approaches in \cite{amlo,nak}, with $a=1$ coincides with the result found in \cite{amlo}. However, the
interesting part is the rotation back in the final result to get a physical Lorentzian
answer. The Wick rotation initially involved setting $\e \rightarrow -\e=1$ to get the
Euclidean set of metrics to be summed over. In the final answer, this rotation is hidden
in the definition of proper-time $T$, or the Euclidean time is $\sqrt\e T$. (This can be
traced back to equation (\ref{act}), and the fact that $\g$ is independent of $\e$). Substituting $T\rightarrow iT$, a physical
Lorentzian answer is obtained. 
 
In two dimensions, we implemented all the properties of Wick rotation stated in section 2, and succeeded in getting an exact and correct answer for the
non-perturbative path-integral. Hence it is interesting to see how far
we can implement this in higher dimensions. 
\subsection{In higher dimensions}

Our formalism crucially rests on the proper-time gauge.
In higher dimensions, as discussed in \cite{dalo},
the action is as given in (\ref{actfix}).

The Euclidean and Lorentzian actions are related by 
$\epsilon\mapsto -\epsilon$. We will set $\epsilon = 1$ in what
follows. 

As discussed in \cite{dalo}, we showed that the above helps in the
identification of the negative definite part of the action, which
is real unlike what a complexification of the Lorentzian metric
would achieve.
\be
S^{\rm ND} =  
 -\frac{(d-1)(d-2)}{16 \pi G}\int d^d x\ {\rm e}^{(d-1) \tilde \lambda}
(\partial_0\tilde \lambda)^2
\ee
with
\begin{equation}
\tilde\lambda = \lambda + \frac{\log \bar g}{2(d-1)}.
\end{equation}
($\l$ is the conformal mode $g_{i j} =e^{2\l}\bar g _{i j}$).
The full action can be written as $S= S^{\rm ND} (\tilde \lambda) + 
S^{\rm rest}(\bar g_{ij}, \tilde{\lambda})$, where we refrain from
giving the remainder $S^{\rm rest}$ explicitly \cite{dalo}. 

The main task is to now deal with the presence of the $e^{\lambda}$
dependence in the measure for $\lambda$. In other words we need
a transformation which is similar to that used in 2-dimensional gravity \cite{dika}:
\be
{\cal D} \tilde g_{ e^{\l} {\bar g}} {\cal D}{\tilde \l}_{e^{\l}{\bar g}}\rightarrow J(\l, {\bar g}) {\cal D}\tilde g_{\bar g}{\cal D}\tilde \l{\bar g}
\ee
We assume that this is achieved with the Jacobian $J$, a local
function of the fields and is of the form $e^{- S_L}$.
For obvious symmetry reasons the local action is of the following form
\be
S_L= \int d^d x \sqrt{\bar g} \le[ A~\le( \p_0 \tilde \l \re)^2 + B~\bar R + C~(\bar R_{i j k l} \bar R^{i j k l}) + D~(\nabla^2 \bar R) + ... \re]
\ee
where A, B, C, D.. are constants whose value will depend on the regularisation
chosen. What is however clear that the kinetic term for the conformal
mode appears in the same form as in the original action and can be absorbed
back into the action with just a renormalisation of the coupling constant
Thus inspired by a conjecture lying at the heart of 2d Liouville quantum 
gravity \cite{dika}, we now assume that an elimination
of the conformal factors from the space-time measures in the 
exponent and the determinants (such that
$\sqrt{\bar g} e^{(d-1) \lambda}\mapsto \sqrt{\bar g}$)
is equivalent to a renormalization of the action. 
We now pull all Jacobians and other $\tilde \lambda$-independent
terms out of the $\tilde \lambda$-integral. The leading divergence
of what is left under the integral is of the form
\begin{equation}
\int {\cal D}{\tilde \lambda} \ e^{\int d^d x \tilde \lambda 
\partial_0^{\dag}\partial_0\tilde \lambda}.
\end{equation}
This formally yields $1/\sqrt{{\det}_{\rm S}(-\partial_0^{\dag}\partial_0)}$,
which is badly divergent, since $\partial_0^{\dag}\partial_0$ is a
positive operator. Nevertheless, it is of exactly the same functional
form as the inverse of scalar part of the Faddeev-Popov determinant as determined
in \cite{dalo}. One can then cancel the two divergent determinants, leaving
behind the integral in $\tilde g_{ij}$. 
This completes the argument for the cancellation of the conformal divergence. 
\section{Discussions}
In this article we argued for a novel type of Wick rotation which
seems particularly suited for a {\it non-perturbative} treatment of
the gravitational path integral. It is an example of how the need
for going beyond perturbative treatments in quantum gravity forces
us to abandon a well-loved, but essentially perturbative concept 
like imaginary time. As an application, we gave a 2dimensional
non-perturbative path integral calculation and 
outlined how in higher dimensions, the conformal sickness of the Wick-rotated path integral is cured non-perturbatively in this new framework.
The two dimensional path-integral revealed that the choice of
proper-time gauge was crucial in our arguements as the Lorentzian
restrictions can be imposed manifestly here. The final result confirmed
that the Wick 
rotation on the space of proper-time metrics does work in a non-perturbative set up. It shall be interesting to see how the Lorentzian conditions obtained here translate to other gauges like the light cone gauge, and this answer reproduced from quantisations used in\cite{ver}. And in higher
dimensions the cancellation of the conformal mode divergence leads to the hope 
that the path-integral can be evaluated to yield exact results.

\noindent
{\bf Acknowledgements}: This paper grew out of extended discussions with R. Loll. The author
wishes to thank her for a careful reading of the manuscript, pointing out of errors,  
criticism of arguments, many helpful comments and encouragement to publish. The author
also thanks B. Mcinnes for bringing to attention typos in eqns 6-8 in earlier version.

\section{Appendix A}
The non-zero Christoffels:
\bea
\G^{0}_{i j} &= &-\frac{\e}{2} {g}_{i j , 0}\\
\G^{i}_{0 j}& = &\frac12 g^{i k} g_{kj , 0}\\
\G^{i}_{l j} &= &\frac12 g^{i k} \le( g_{ l k, j} + g_{j k, l} - g_{l j, k}\re)
= \G^{(2)\ i}_{\ \ lj }
\eea
The Ricci Tensor components
\bea
R_{00} &= &\frac12 \le(g^{i k} g_{ i k, 0}\re)_{, 0} + \frac14 g^{ jm} g^{kn} g_{jn,0} g_{k m, 0}\\
R_{i 0} &=& \G^{(2) j}_{\ \ i j, 0}  -\frac12 \le(g^{j k} g_{ i k, 0}\re), j 
+ \G^{ (2) k}_{ i j} \frac12 g^{j m} g_{k m, 0} - \G ^{(2) j}_{jk} \frac12 g^{k l} g_{ il, 0}\\
R_{ ij } &= & R^{ (2)}_{ i j} + \e \le( g_{i j, 0, 0} + g_{ij, 0} g^{ lm} g_{lm, 0}
- g_{i k, 0} g_{ j m, 0} g^{k m}\re)
\eea

Scaling the time coordinate by $\sqrt{\e}$, and putting the above in
Einstein's equation, the entire $\e$ dependence is removed from them
(we also need to use $\e^2=1$). So, one obtains $g_{ij}= g_{ij}(\sqrt{\e}t,x)$

\section{ Appendix B}
Though the 2 dimensional classical action is a total derivative, one knows 
that dynamics can be induced in the quantum corrected effective action
due to the conformal anamoly. In anticipation of that, one can
try and see what the classical reduced space can be, by looking
at the constraint $T_{01}=0$. This can be imposed by using
\be
\frac{\delta S}{\delta g^{01}}|_{g_{01}=0}= 0
\label{const3}
\ee
with $S= \int \sqrt{g} R + \Lambda \sqrt g$. Thus using the non-zero Christoffels, one finds
\be
S= \int {\sqrt \g}\left[ \e R_{00} + \frac1{\g} R_{11} \right] + O(g_{01}^2)
\ee

With(quoting only the terms proportional to $g_{01}$) and $\e =1$,
\bea
R_{00}&=& \partial_x(\frac1{\g} \partial_t g_{10})  + \frac1{2 \g^2}\partial_t g_{10} \partial_x\g\\
R_{11} &=& \partial_x\partial_t g_{01} - \frac{1}{4 \g^2}g_{01}\partial_t \g\partial_x\g 
- \frac{1}{2\g}\left( \partial_t\left(\frac{g_{01}}\g\right)\partial_x \g - \partial_t\g \partial_x \left(\frac{g_{01}}{\g}\right)\right)
\eea
Using the above in (\ref{const3}), one finds that 

\be
T_{01} = \partial_x\partial_t \ln \sqrt \g =0
\ee
is solved by $\g= \g(t)\g(x)$

Thus classical configuration space is reduced by the above constraint.
We can try to see what this implies for the conformal gauge:
The coordinate transformations relating metric components in
conformal gauge (\ref{conm}) and (\ref{ptm}). 
\bea
\e~ \dot t^2 + \g ~\dot x^2 &= & \e~ e^{2\phi} \label{tr1}\\
\e~ \dot t t' + \g ~\dot x x'& = & 0 \label{tr2}\\
\e~ t'^2 +  \g x'^2 =  e^{2\phi} \label{tr3}
\eea

Dividing (\ref{tr1}) by (\ref{tr3}) and using (\ref{tr2})
one obtains:
\bea
\dot t &= &\sqrt{\g} x'\\
 t' & = &\sqrt{\g} \dot x
\eea

Putting the above two in (\ref{tr1}) and (\ref{tr3}) one obtains
(\ref{trans1},\ref{trans2}).

To follow on the heuristic arguement, and Equation (\ref{cond}), we show 
why the condition $<T_{00}>\neq \frac1{\g} <T_{11}>$.
The conformal anamoly equation, and the conservation equations
$\nabla^{\mu}<T_{\mu \nu}> = 0$ can be used to solve for the
energy momentum tensors. For clarity we denote the the conformal
anamoly in this gauge as $<T_{00}> + \frac{1}{\g} <T_{11}>= F(\g)$,
where $F(\g)$ is an arbitrary function of the metric, and is usually
proportional to the scalar curvature. In the next few lines $<T_{\m \n}>\equiv
T_{\m \n}$.
The $T_{01}$ component of the energy momentum tensor can be eliminated,
yielding the following two equations:
\be
\partial_1 \left\{\frac{1}{\sqrt \g}\partial_1\left(\frac{1}{\sqrt \g} T_{11}\right)\right\} + \partial_0\left\{\sqrt{\g} \partial_0 \left(\frac{T_{11}}{\sqrt \g}\right)\right\} 
+ \partial_0\left\{\dot {\ln \g} T_{11}\right\} = \partial_0\left\{\sqrt{\g} \partial_0( \sqrt \g F(\g))\right\}
\ee
\be
\partial_0\left\{\sqrt\g \partial_0 \left(\sqrt \g T_{00}\right)\right\} + \partial_1\left\{\frac1{\sqrt \g}\partial_1\left(\sqrt\g T_{00}\right)\right\} = -\partial_1\left\{\frac1{\sqrt \g}\partial_1\left(\sqrt{\g} F(\g)\right) \right\} - \partial_0\left\{\g \dot{\ln  \g} F(\g)\right\}.
\ee 

The above two equations can be solved in principle to obtain a solution for the
energy momentum tensor components, however it is obvious that $T_{00}= \frac{1}{\g}T_{11}$ cannot be a solution of the above. Hence Equation (\ref{cond}) are 
disallowed.
\section{Appendix C}

This is a summary of the Faddeev-Popov determinant calculations in our gauge.
An ab initio derivation of the determinant is required as our gauge is {\it non-covariant}
and two dimensions is the easiest place to demonstrate that.

By ab-initio we mean that we start with the Gaussian-Normalisation condition:
\be
\int {\cal D}\d g_{\mu \nu} e^{-~<\delta g,\delta g>} = 1
\label{gn}
\ee

where $\delta g_{\mu \nu}$ is fluctuation in tangent space
around a given base point in the functional space of metrics.

We assume that our gauge fixed proper-time metric, is of the following form:
\be
g_{\m \n}= \left(\begin{array}{cc}1&0\\0&\g\end{array}\right)
\ee
with the following non-zero Christoffel symbols.
$$\G^{0}_{11}= -\frac{1}{2}\dot{\g} , \ \  \G^{1}_{01}= \frac{1}{2}\dot{(\ln {\g})} , \ \ \G^{1}_{11} = \frac{1}{2}
(\ln \g)'$$
$\dot \g \equiv \partial_t\g, ~\g' \equiv \partial_x \g$ denote the derivatives and $\g= \g(t)\g(x)$.

Using the above: 

\bea
<\delta g,\delta g> &=& \int d^2 x \sqrt {\g} \left[ ( 2\nabla_0\xi_0)^2  + \frac2{\g}(\nabla_0\xi_1
+\nabla_1\xi_0)^2 + \frac1{\g^2}(\d g_{11} + 2\nabla_1\xi_1)^2\right] \nn \\
&=& \int d^2 x \sqrt{\g} \left[   \frac{(\d g_{1 1})^2}{\g^2} + 4 {\dot \xi_0}^2 +  \right. \nn \\ &+& \frac4{\g^2}\left( \left\{(\xi_1') + \frac{1}{2}\dot{\g} ~\xi_0 - \frac12(\ln \g)' ~\xi_1\right\}^2  + 
 \left\{ \xi_1' + \frac12 \dot{\g} ~\xi_0 - \frac12 (\ln \g)' \xi_1\right\}\delta g_{11}\right) \nn \\& +&  \left. \frac2{\g}\left( \dot \xi_1 + \xi_0' - \dot{(\ln \g)} \x_1\right)^2\right]
\eea

Using partial integrations,  the above can be written in a more convenient form:
\bea
&= &\int d^2x \sqrt{\g} \left[ \frac{1}{\g^2}\d g_{11 }^2 + 4 \xi_0\left(\p_0^{\dag}\p_0 + \frac{1}{2\g}\p_1^{\dag}\p_1 + \frac{{\dot{(\ln \g)}}^2}{4}\right)\xi_0 \right. \nn \\
&+&\frac{4}{\g}\left\{\xi_0 \dot{(\ln \g)}(\p_1 - \frac12 \ln \g')\xi_1\right\} - \frac4{\g}\left\{
\xi_1\left(\p_0 + \frac12 \dot{(\ln \g)}\right)\p_1 \x_0 \right\} \nn\\
&+& \frac{4}{\g^2}\left\{\xi_1\left(\p_1^{\dag}\p_1 + \frac{\g}{2}\p_0^{\dag}\p_0 + \frac{\g\dot{\ln \g}}{2}\p_0 + \frac{(\ln \g')^2}{4}\right)\x_1\right\} \nn\\
&+& \left.  \frac{4}{\g^2}\xi_1\p_1^{\dag}\d g_{11} + \frac2{\g}{\dot {(\ln \g)}}\xi_0\d g_{11}\right]
\eea

(where $-\partial_0^{\dag}= \p_0 + 1/2 \dot{(\ln \g)},\ \ \  -\p_1^{\dag} = \p_1 - 1/2 (\ln \g)'$)

One can then complete the squares to do the $\xi$ integrations and obtain:
\be
=\int d^2 x \frac{1}{\g^2} \d g_{11}^2 + \xi'_0 \nabla_0\xi'_0 + \x'_1\nabla_1\x'_1 - W_0 - W_1
\label{det}
\ee

where, 
\bea
\nabla_0 &= & 4(\p_0^{\dag}\p_0 + \frac1{2\g}\p_1^{\dag}\p_1 + \frac{{\dot{(\ln \g)}}^2}{4})\\
\nabla_1&=&  \frac{4}{\g^2}(\p_1^{\dag}\p_1 + \frac{\g}{2}\p_0^{\dag}\p_0 + \frac{\g}{2}\dot{(\ln \g)}\p_0 + \frac{(\ln\g')^2}{4}) \nn \\ 
&+& \frac{1}{\g}(\p_0^{\dag} - \dot{\ln \g})\p_1 (\nabla_0)^{-1}\frac{1}{\g}\p_1^{\dag}(\p_0 - 2 \dot{\ln \g}) \label{ope11}\\
W_0 &=& \frac1{\g} \d g_{11} \dot{\ln \g}\nabla_0^{-1}\frac{\dot{\ln \g}}{\g}\d g_{11}\\
W_1&=& \d g_{11}\frac1{\g}\le(\p_1 + \dot{\ln \g}\p_1^{\dag}(\p_0 - 2\dot{\ln \g})\p_1 \nabla_0^{-1}\re)\nabla_1^{-1} \nn \\
&\times& \le( (\p_0^{\dag} - \dot{\ln \g})\p_1\nabla_0^{-1}\frac{\dot{\ln \g}}{\g} + \p_1^{\dag}\re)\d g_{11} \label{ope22}
\eea

After this, one can just evaluate the gaussian integrals in $\d g_{11}$,
$\xi_0,\xi_1$ in the (\ref{gn}) to get the Jacobian in terms of three scalar determinants,\\ $det_S(\nabla_0) det_S(\nabla_1) det_S(\nabla_{11})$, with $\d g_{11}(\nabla_{11} -1/\g^2)\d g_{11} =  - W_0 -W_1$.
For non-zero $\g$ it becomes a Herculian task to determine the exact form of the
effective action. Following the zeta function regularisation methods
used in \cite{ram}, the form of the effective action can be guessed
by examining around a value of the metric for which $\dot{\ln \g}=0$.
This gives precisely the same action found in \cite{nak} for the determinant
$\nabla_0$ as explained in section 3.1. For the rest of the determinants
of operators (\ref{ope11}) and (\ref{ope22}), we also do a zeta function
regularisation. But most give diverging results due to presence of inverse
of operators, and we absorb these in an infinite renormalisation of the
effective action.

\end{document}